# Permanent matching of coupled optical bottle resonators with better than 0.16 GHz precision

N. A. T‌OROPOV[1], AND M. S‌UMETSKY[2,*]

[1]ITMO University, St.Petersburg, 197101, Russia
[2]Aston Institute of Photonic Technologies, Aston University, Birmingham B4 7ET, UK,
*Corresponding author: m.sumetsky@aston.ac.uk

**The fabrication precision is one of the most critical challenges on the way to the creation of practical photonic circuits composed of coupled high Q-factor microresonators. While very accurate transient tuning of microresonators based on local heating has been reported, the record precision of permanent resonance positioning achieved by post-processing is still within 1-5 GHz. Here we demonstrate two coupled bottle microresonators fabricated at the fiber surface which resonances are matched with a better than 0.16 GHz precision. This corresponds to a better than 0.17 angstrom precision in the effective fiber radius variation. The achieved fabrication precision is only limited by the resolution of our optical spectrum analyzer and can be potentially improved by an order of magnitude.**

Fabrication precision is one of the most important challenges on the way to a practical microphotonic technology for the future optical communications, quantum computing, microwave photonics, and ultraprecise optical measurement devices. The outstanding fabrication precision achieved in silicon photonics [1-3] is still far beyond the requirements for the creation of practical miniature slow light delay lines, buffers, and other optical signal processing devices [4-6]. Similarly, improvement of the fabrication precision is critical for the prospective microwave photonics, which requires optical filters having a flat top spectrum within very narrow bandwidth and, simultaneously, exceptionally high rejection rate [7, 8]. As another example, the bandwidth and repetition rate of microresonator comb generators is determined by their dispersion [9], which can be controlled by the prospective ultraprecise fabrication technology of these resonators.

All these and many other potential applications of modern microphotonics rely on the success of ultraprecise fabrication of individual microresonators and microresonator photonic circuits. The ultrahigh Q-factors of these microresonators ranging from $Q=10^6$ in silicon photonics [2, 3] to $Q=10^9$ for the whispering gallery mode (WGM) microresonators [10, 11] have been demonstrated. In order to arrive at the best performance of the microphotonic devices mentioned above, the positioning precision of resonances are anticipated to match or be better than the resonance width, which ranges from ~ 0.1 GHz for $Q=10^6$ to ~ 0.1 MHz for $Q=10^9$ at 1.5 μm radiation wavelength. Very accurate post-processing of individual spherical and toroidal microresonators by chemical etching with the resonance positioning precision of 0.1 GHz and MHz-scale control of their free spectral range has been demonstrated [11, 10]. However, this approach is not applicable to the local post-processing of photonic circuit elements. While very accurate transient tuning based on local heating of coupled microresonators has been reported [2, 3, 13, 14], the record precision of permanent resonance positioning for these structures achieved by post-processing was within 1-5 GHz [15-18]. Thus, the precise fabrication of microresonator circuits with a predetermined resonance positioning still remains a major challenge on the way of creation of practical microresonator photonic circuits.

Recently a novel photonic fabrication platform called Surface Nanoscale Axial Photonics (SNAP) demonstrated fabrication of miniature WGM resonant photonic circuits at the surface of an optical fiber with unprecedented sub-angstrom precision [19-21]. Photonic circuits are introduced in SNAP by nanoscale effective radius variation (ERV) of the optical fiber using a focused $CO_2$ laser beam. In [20] a structure of 30 coupled resonators was fabricated with the 0.7 angstrom precision. In [21], a breakthrough SNAP bottle resonator miniature delay line was fabricated with the precision of 0.9 angstrom. In [22], it was shown that fully reconfigurable SNAP structures can be introduced at the fiber surface by local heating and translated with sub-angstrom precision. However, further significant improvement in the precision of SNAP is required for the fabrication of practical miniature optical buffers and signal processing devices [6].

Here we advance the fabrication precision by a factor of 4 compared with that achieved in SNAP [20] and by an order of magnitude compared with the precision of resonance frequency positioning demonstrated for planar resonant circuits [15-18]. We demonstrate a better than 0.16 GHz precision in the positioning of a resonance, which corresponds to a better than 0.17 angstrom precision in the ERV of a microresonator, limited by the resolution of the optical spectral analyzer used. This result paves the way to the ultraprecise fabrication of practical miniature delay lines, signal processing devices, microwave photonics filters, and other microphotonic devices for the future optical communications and ultraprecise optical measurements.

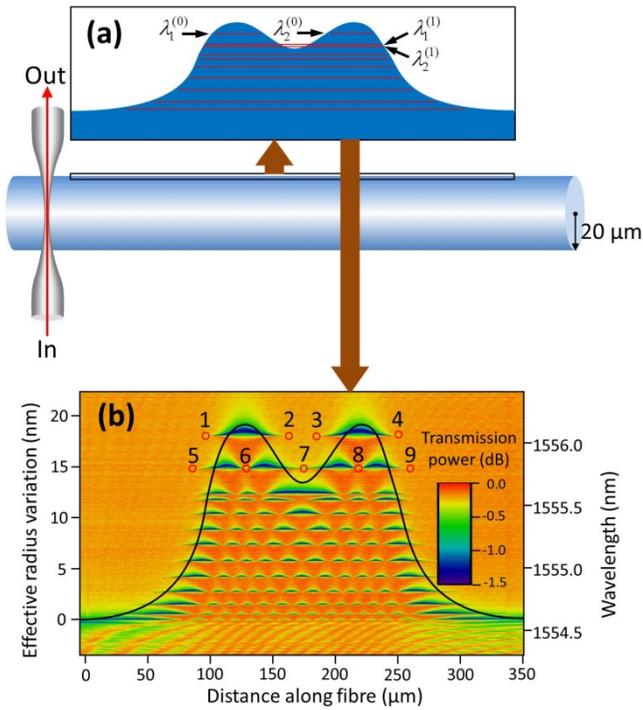

Fig. 1. (a) – Illustration of a fiber with introduced two nanoscale-high coupled bottle resonators (inset). The WGMs in the resonators are excited by a transverse fiber taper. (b) – Surface plot of the resonant transmission power spectrum measured with 2 μm resolution along the fiber axis. Black curve is the ERV of the fabricated coupled bottle resonators. The resonances near points 1-9 are shown in Fig. 2(a) and (b).

Fig. 1(a) illustrates our experiment. To test the fabrication precision, two coupled bottle resonators having nanometer scale ERV are introduced at the surface of an optical fiber by annealing with a focused $CO_2$ laser beam. The eigenvalues of this structure are enumerated by three quantum numbers $(m, n, p)$, where $m$ is the azimuthal, $n$ is the radial and $p$ is the axial number. In the experiment below, the fiber radius is 20 μm. In the C-band considered, the spacing between the adjacent azimuthal resonances (corresponding to $m$ and $m+1$) is approximately 15 nm, the spacing between the adjacent radial resonances is several nanometers, while the spacing between the adjacent axial resonances is much smaller and can be controlled by the nanoscale ERV of the fiber. Fig. 1(a) shows the axial resonances of the double bottle resonator (red lines) which correspond to a fixed $m$ and $n$ quantum numbers and different $p$.

To test the fabrication precision, one of resonators is used as a reference and the other one is post-processed in order to minimize the difference between the reference and adjusted resonances. In our experiment we go further. We chose the distance between two bottle resonators to be large enough so that the splitting between their fundamental axial resonances $\lambda_1^{(0)}$ and $\lambda_2^{(0)}$ corresponding to $p=0$ (inset in Fig. 1(a)) is unresolved within the precision of measurements. At the same time, we chose this distance to be small enough to resolve the splitting between the next resonances $\lambda_1^{(1)}$ and $\lambda_2^{(1)}$ corresponding to $p=1$. The axial distribution of WGMs localized in a SNAP resonator is described by a one-dimensional Schrödinger equation with potential $V(z)$ proportional to the ERV $\Delta r_{eff}(z)$ [19]. Therefore, the relation between the shift of resonances in coupled resonators, $\lambda_k^{(1)} - \lambda_k^{(2)}$, and the shift of resonances in the independent resonators, $\lambda_{0k}^{(1)} - \lambda_{0k}^{(2)}$, is found from [23]

$$|\lambda_k^{(1)} - \lambda_k^{(2)}| = \sqrt{(\lambda_{0k}^{(1)} - \lambda_{0k}^{(2)})^2 + \Omega_k^2}. \qquad (1)$$

where $\Omega$ determines the coupling between resonances. From this equation, the measured shift of resonances always exceeds their unperturbed shift, which determines the actual axial asymmetry of $\Delta r_{eff}(z)$.

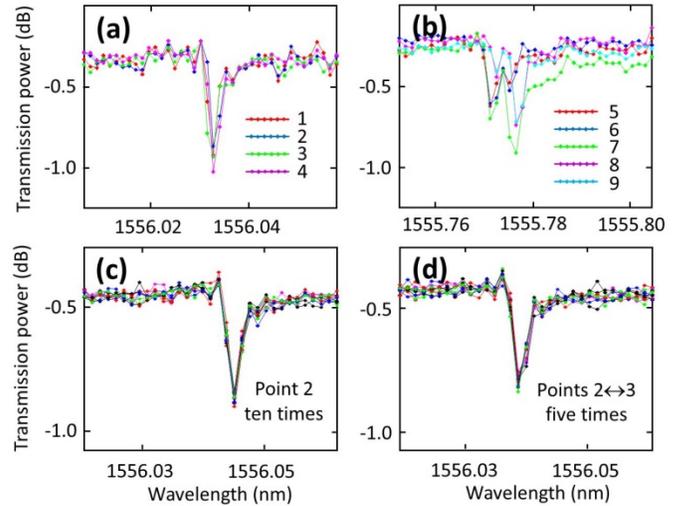

Fig. 2. (a) – $p=0$ resonance measured near points 1, 2, 3, and 4 of Fig. 1(b). (b) – split $p=1$ resonance measured at points 5, 6, 7, 8, and 9. (c) – $p=0$ resonance periodically measured at point 2 ten times during 5 minutes. (d) $p=0$ resonance periodically measured at points 2 and 3, switched 5 times during 3 minutes.

The spectrum of the introduced structure is measured by a Luna optical spectrum analyzer (OSA) connected to the transverse biconical fiber taper with a micron waist (Fig. 1(a)). The waist periodically touches the fiber in the process of measurement. As noted, we set the separation between the bottle resonators along the fiber axis so that the splitting between their axial fundamental modes $\Omega_0$ is much smaller than the measurement resolution, while the splitting between $p=1$ modes $\Omega_1$ is the smallest possible yet resolved by the OSA (0.16 GHz resolution corresponding to 1.3 pm at 1.55 μm wavelength). The $p=0$ resonances of the originally introduced bottles deviated by a few GHz due to minor fabrication errors. In order to equalize them, we performed post-processing. We calibrated the $CO_2$ laser beam power and exposure time so that a single laser shot introduced a shift of a $p=0$ resonance which is smaller than the measurement resolution 0.16 GHz. Next, the number of shots, which was necessary to match the resonances, was calculated and the shots were introduced into one of the bottles. The post-processing ensured that the $p=0$ resonances coincide with the precision better than the measurement resolution 0.16 GHz, because each of laser shots introduced the resonance shift which was smaller than the measurement resolution.

The post-processed double bottle resonator structure was characterized by scanning the taper along the fiber axis (Fig. 1(a)) and measuring the resonance spectrum with the 2 μm intervals. The results of measurements are presented by the surface plot in Fig. 1(b), which determines the transmission power as a function of axial position of the taper and wavelength. In this plot, the ERV $\Delta r$ (right axis) is rescaled from the wavelength variation $\Delta \lambda$ (left axis) using relation [12, 19]

$$\Delta r = \frac{\Delta \lambda}{\lambda_0} \cdot r_0 ,\qquad(2)$$

where the fiber radius $r_0 = 20$ nm and wavelength $\lambda_0 = 1550$ nm. Fig. 2(a) compares the $p=0$ resonance spectrum near left bottle (points 1 and 2 in Fig. 1(b)) and right bottle (points 3 and 4 in Fig. 1(b)) and confirms that these resonances coincide to within the resolution of measurements. Fig. 2(b) compares the split $p=1$ resonance along the length of the double bottle structure (points 5, 6, 7, 8, and 9 in Fig. 1(b)) showing remarkable reproducibility of the resonance positions and their splitting ~ 0.5 GHz. To the best of our knowledge, the latter value is the smallest permanently introduced splitting of resonances reported for coupled microresonators.

To double check the achieved fabrication precision we verified the reproducibility of spectral measurements and also excluded the effect of temperature variation. Fig. 2(c) shows 10 periodic measurements of the resonance at point 2 in Fig. 1(b) acquired during 5 minutes, which confirm the stability of measurements to within the resolution of the OSA used. Fig. 2(d) shows 10 measurements of resonances (5 at point 2 of the left bottle and 5 at point 3 of the right bottle, periodically switched) acquired during 3 minutes indicating that the position of these resonances coincide to within the measurement resolution. Overall, we have confirmed that the $p=0$ resonances of the bottle resonators coincide with the precision better than the OSA resolution equal to 0.16 GHz.

Finally, the deviation in ERV of the bottle resonators $\delta r$ is expressed through the deviation in the positions of their $p=0$ resonances $\delta \lambda$ by Eq. (2), $\delta r = r_0 \cdot \delta \lambda / \lambda_0$, [12, 19] rescaling left and right axes in Fig. 1(b). From here, the achieved precision of ERV is found as better than 0.17 angstrom. More accurate determination of the fabrication precision can be performed by solving the inverse problem for the Schrödinger equation, which determines the spectrum of a bottle resonator with a given ERV. Solution of this problem is beyond the scope of this Letter.

In summary, we have demonstrated the fabrication of coupled bottle resonators with a better than 0.16 GHz precision in positioning of resonances of the bottle resonators which corresponds to a better than 0.17 angstrom precision in the effective radius variation of an optical fiber. This is a factor of 4 improvement of the result previously achieved in SNAP [20] and an order of magnitude improvement of the precision achieved in planar photonics technologies [15-18]. While this demonstration was concerned with a simplest structure of two coupled resonators, the nature of the iterative post-processing technique used suggests that more complex photonics circuits can be fabricated similarly. Furthermore, since our measurements were limited by the resolution of the optical spectrum analyzer used, the demonstrated fabrication precision can be advanced with more precise spectrum measurements and further optimization of the $CO_2$ laser beam power and exposure time. It is anticipated that the future development of the SNAP technology based on this demonstration will pave the way to the creation of practical miniature photonic circuits for applications ranging from optical communications and quantum computing to ultraprecise time and frequency measurement technologies.

**Funding.** Royal Society (WM130110). Horizon 2020 (H2020-EU.1.3.3, 691011); British Council Researcher Links Travel Grant (164861753).

**Acknowledgments**. MS acknowledges the Royal Society Wolfson Research Merit Award.